\documentclass[conference]{IEEEtran}
\ifCLASSINFOpdf
  \usepackage[pdftex]{graphicx}
  \usepackage[utf8]{inputenc}
  \usepackage{algorithm2e}
  \usepackage{algpseudocode}
  \usepackage{algcompatible}
   \usepackage{subfigure}
   \usepackage{cite}
  \RestyleAlgo{ruled}
\else
\fi
\usepackage{amsmath,amsfonts}
\usepackage{xcolor}

\usepackage[top=0.75in,bottom=1in,left=1in,right=1in]{geometry}

\begin{document}
%
\title{Dynamic Coded Distributed Convolution for UAV-based Networked Airborne Computing}



%
\author{\IEEEauthorblockN{Bingnan Zhou\IEEEauthorrefmark{1}\IEEEauthorrefmark{2},
Junfei Xie\IEEEauthorrefmark{1},
Baoqian Wang\IEEEauthorrefmark{1}\IEEEauthorrefmark{2}}
\IEEEauthorblockA{\IEEEauthorrefmark{1} Department of Electrical and Computer Engineering\\
San Diego State University,
San Diego, CA, 92182}
\IEEEauthorblockA{\IEEEauthorrefmark{2}
University of California San Diego, San Diego, CA, 92093\\
Email: jxie4@sdsu.edu
}
}


\maketitle


\begin{abstract}
A single unmanned aerial vehicle (UAV) has limited computing resources and battery capacity, making it difficult to handle computationally intensive tasks such as the convolution operations in many deep learning applications. UAV-based networked airborne computing (NAC) is a promising technique to address this challenge. It allows UAVs within a range to share resources among each other via UAV-to-UAV communication links and carry out computation-intensive tasks in a collaborative manner. This paper investigates the vector convolution problem over the NAC architecture. 
A novel dynamic coded convolution strategy with privacy awareness is developed to address the unique features of UAV-based NAC, including node heterogeneity, frequently changing network typologies, time-varying communication and computation resources. Simulation results show its high efficiency and resilience to uncertain stragglers. 
\end{abstract}


%
\IEEEpeerreviewmaketitle

\section{Introduction}
Recent years have witnessed the fast popularization of unmanned aerial vehicle (UAV) in both academia and industry \cite{honkavaara2013processing, choi2009developing, ham2016visual}. The UAVs are often equipped with sensing, communication, and computing capabilities and can generate massive data, which include valuable information that can be used for improving decisions, making scientific discoveries, or supporting new artificial intelligence (AI) applications. To effectively utilize the information from these data, e.g., by using deep learning algorithms, considerable amount of computing resources are often needed. However, due to small payload, a single UAV often has a limited computing capability  and  battery capacity  for carrying out computation-intensive tasks. 

To address above issues, the existing solution is to offload data from the UAV to the ground station or remote cloud for processing. 
However, this solution suffers from many issues such as long transmission latency and data losses, and is thus not suitable for delay-sensitive applications. A better solution is to offload the data to nearby UAVs and leverage their computing resources to perform data processing and analysis. Such a computing system formed by UAVs  connected via UAV-to-UAV communication links are often known as the UAV-based networked airborne computing (NAC) \cite{lu2019toward}. Compared with the traditional cloud- or static server-based computing systems, UAV-based NAC systems are featured by 1) high node mobility;  
2) heterogeneous nodes with different computing, communication and sensing capabilities; and 3) dynamic computing and communication resources. These unique features make many existing distributed computing techniques that assume homogeneous and static computing nodes perform poorly in UAV-based NAC systems. 

The time-varying communication and computing properties of UAV-based networked airborne computing systems can be modeled as uncertain stragglers that are slow in generating the result or take a long time to transmit data. Topology changes or link/node failures can also be modeled as uncertain stragglers that fail to generate or return any results. To alleviate the effects of stragglers, 
coded distributed computing (CDC) \cite{ng2021comprehensive} is a promising technique, which introduces computation redundancy into the system via exploiting the coding theory. Currently, most works on CDC focus on the matrix multiplication problem or assume  homogeneous distributed systems with static computing nodes \cite{lee17coded,wang2021batch,wang2019coding,reisizadeh17coded}. However, many data analysis algorithms, especially the filtering or feature extraction techniques like the convolutional neural networks (CNNs), involve convolution operations. How to perform resilient distributed convolution over UAV-based NAC systems formed by heterogeneous moving UAVs has not been investigated, to the best of our knowledge. The state-of-the-art coded convolution strategy introduced in \cite{dutta2017coded} was designed for homogeneous systems with static computing nodes, which performs poorly over the UAV-based NAC system as we will show in the simulation studies. Although there have been some works considering heterogeneous systems \cite{reisizadeh17coded,wang2021batch,wang2019coding} and moving computing nodes \cite{keshtkarjahromi2018dynamic,wang2021multi}, these works are centered on the matrix multiplication problem, which has a quite different problem solving procedure from  the convolution problem. 

In this paper, we aim to fill the aforementioned research gap by making the following main contributions: 
\begin{itemize}
\item  \textit{Dynamic coded distributed convolution strategy}. We propose an innovative dynamic coded distributed convolution strategy with privacy awareness for UAV-based NAC.  It integrates the coding theory with a novel task decomposing and allocation mechanism to dynamically assign tasks to the worker nodes based on their communication and computing performances. Unlike most existing CDC algorithms that have to pre-determine the amount of computation redundancy to be introduced before performing the task, our strategy introduces redundancy dynamically and only when needed. It can thus achieve high resilience with the minimal redundancy.    
Furthermore, as our strategy encodes the input data, data privacy is protected to some extent.
\item \textit{Comprehensive simulation studies.} We conducted comprehensive simulation studies  to evaluate the performance of the proposed strategy, in comparison to the uncoded distributed convolution strategy and the state-of-the-art coded distributed convolution strategies. The results demonstrate the high efficiency and resilience of the proposed strategy in face of uncertain stragglers. 
\end{itemize}

In the rest of the paper, we first describe the problem to be solved in Sec. \ref{sec:problem_desc}, and then review the two existing distributed convolution strategies in Sec. \ref{sec:review}. The proposed dynamic coded distributed convolution strategy is then introduced in Sec. \ref{sec:method}. In Sec. \ref{sec:simulation}, we present the simulation results on the performance of the proposed strategy, compared to existing distributed convolution strategies. Section \ref{sec:conclusion} finally concludes the paper. 

\section{Problem Description}
\label{sec:problem_desc}
Consider a UAV-based NAC system formed by multiple UAVs with different computing and/or communication capabilities. Suppose one of the UAV needs to perform a vector convolution task, $\boldsymbol{a}*\boldsymbol{x}$, where $\boldsymbol{a}\in \mathbb{R}^{N_1}$ is a pre-stored vector and $\boldsymbol{x} \in \mathbb{R}^{N_2}$ is the input vector. To save energy and reduce computation time, it decides to offload the task to its neighbors within its communication range. 

The problem considered in this paper is how the \textit{master node} (UAV that offloads the task) should decompose the task and distribute subtasks to surrounding \textit{worker nodes} (UAVs that execute the offloaded task collaboratively), such that the task completion time is minimized. To solve this problem, the key technical challenges to conquer include: 1) As all worker nodes in the UAV-based NAC system can move, the network topology may change frequently due to node leave and join, and the communication quality of UAV-to-UAV links varies over time; 2) The computing resources available at a worker node are also time variant, due to completion of old tasks or receipt of new tasks; 3) The input data $\boldsymbol{x}$ may contain sensitive information and directly sending the data to worker nodes may raise privacy concerns. The desired distributed computing scheme should thus 1) be resilient to network topology and resource changes, 2) be efficient in computing the task, and 3) protect data privacy to certain extent. 


\section{Review of Existing Solutions}
\label{sec:review}
In this section, we review two state-of-the-art distributed convolution strategies. 
\subsection{Uncoded Convolution Strategy} 
In the uncoded convolution strategy introduced in \cite{dutta2017coded}, the master node first partitions both vectors $\boldsymbol{a}$ and $\boldsymbol{x}$ evenly into a set of sub-vectors of length 
$s=\sqrt{\frac{N_{1}N_{2}}{P}}$,  i.e., $\{\boldsymbol{a}_{1}, \boldsymbol{a}_{2}, ..., \boldsymbol{a}_{\frac{N_{1}}{s}}\}$ and $\{\boldsymbol{x}_{1}, \boldsymbol{x}_{2}, ..., \boldsymbol{x}_{\frac{N_{2}}{s}}\}$, where $P$ is the total number of worker nodes. It then sends each pair of sub-vectors, $\boldsymbol{a}_i$ and $\boldsymbol{x}_j$, to a different worker node for further processing, 
where $1 \leq i \leq \frac{N_{1}}{s}$ and $1 \leq j \leq \frac{N_{2}}{s}$. Each worker node computes $\boldsymbol{a}_i*\boldsymbol{x}_j$ and returns the result back to the master node. After receiving results from all worker nodes, the master node finally aggregates the results with proper shifts to obtain the value of $\boldsymbol{a}*\boldsymbol{x}$. 

As this strategy requires the results from all worker nodes to obtain the final value, any delay  will significantly degrade its performance and any node/link failure will cause the whole task to fail. In addition, this strategy simply decomposes the workload evenly, and thus cannot address the node heterogeneity and dynamic features of UAV-based NAC. Moreover, it directly sends the input data to the worker nodes and hence may cause information leakage.  

\subsection{Traditional Coded Convolution Strategy}
\label{coded_method}
To improve the resilience of the uncoded strategy to the straggler effects, a coded strategy was developed in \cite{dutta2017coded}. The key idea is to introduce redundancy into the computation by using the coding theory. In particular, similar to the uncoded strategy, the coded strategy first partitions both vectors $\boldsymbol{a}$ and $\boldsymbol{x}$ into small sub-vectors of equal length $s$. The difference is that the sub-vector length $s$ can be any value larger than $\sqrt{\frac{N_{1}N_{2}}{P}}$, and the $\frac{N_{1}}{s}$ sub-vectors of $\boldsymbol{a}$ are encoded into $\frac{Ps}{N_{2}}$ sub-vectors with each having a length of $s$, by using a ($\frac{Ps}{N_{2}}$, $\frac{N_{1}}{s}$) MDS code. 
The following equation shows how a Vandermonde matrix, denoted as $V \in \mathbb{R}^{\frac{N_{1}}{s} \times \frac{Ps}{N_{2}}}$, can be used to encode the set of $\frac{N_1}{s}$ sub-vectors,  $\{\boldsymbol{a}_1,\boldsymbol{a}_2,...,\boldsymbol{a}_{\frac{N_1}{s}}\}$, into a larger set of $\frac{Ps}{N_2}$ sub-vectors, $\{\boldsymbol{\hat{a}}_1,\boldsymbol{\hat{a}}_2,...,\boldsymbol{\hat{a}}_{\frac{Ps}{N_2}}\}$:
\begin{align*}
\begin{bmatrix}
  \hat{\boldsymbol{a}}_{1} \\
  \hat{\boldsymbol{a}}_{2} \\
  \vdots \\
  \hat{\boldsymbol{a}}_{\frac{Ps}{N_{2}}}
\end{bmatrix} &=
V \begin{bmatrix}
  \boldsymbol{a}_{1} \\
  \boldsymbol{a}_{2} \\
  \vdots \\
  \boldsymbol{a}_{\frac{N_{1}}{s}} \\
\end{bmatrix} \nonumber\\
&= \begin{bmatrix} 
  1 & g_{1} & g_{1}^{2} & \dots & g_{1}^{\frac{N_{1}}{s} - 1} \\
  1 & g_{2} & g_{2}^{2} & \dots & g_{2}^{\frac{N_{1}}{s} - 1} \\
  \vdots & \vdots & \vdots & \ddots & \vdots \\
  1 & g_{\frac{Ps}{N_{2}}} & g_{\frac{Ps}{N_{2}}}^{2} & \dots & g_{\frac{Ps}{N_{2}}}^{\frac{N_{1}}{s} - 1} 
\end{bmatrix}
\begin{bmatrix}
  \boldsymbol{a}_{1} \\
  \boldsymbol{a}_{2} \\
  \vdots \\
  \boldsymbol{a}_{\frac{N_{1}}{s}}
\end{bmatrix} \nonumber 
\end{align*}

With the encoded sub-vectors $\{\hat{\boldsymbol{a}}_i\}^{\frac{Ps}{N_2}}_{i=1}$, the master node then sends each pair  $(\hat{\boldsymbol{a}}_i,  {\boldsymbol{x}}_j)$ to a different worker node, where $1 \leq i \leq \frac{Ps}{N_{2}}$ and $1 \leq j \leq \frac{N_{2}}{s}$. The worker nodes then convolve the received two sub-vectors and return the result back to the master node after the task is completed. The master node can decode $\{{\boldsymbol{a}}_i * {\boldsymbol{x}}_j|1 \leq i \leq \frac{N_1}{s}\}$ to reconstruct $\boldsymbol{a}*\boldsymbol{x}_{j}$ after receiving any $\frac{N_1}{s}$ of the set  $\{\hat{\boldsymbol{a}}_i * {\boldsymbol{x}}_j|1 \leq i \leq \frac{Ps}{N_{2}}\}$ by using the following equation: 

\begin{align*}
\begin{bmatrix}
  {\boldsymbol{a}}_1 * {\boldsymbol{x}}_j  \\
  {\boldsymbol{a}}_2 * {\boldsymbol{x}}_j  \\
  \vdots \\
  {\boldsymbol{a}}_{\frac{N_{1}}{s}} * {\boldsymbol{x}}_{j}  \\
\end{bmatrix}
&= V_{j}^{-1} \begin{bmatrix}
  \hat{\boldsymbol{a}}_{i_{1}} * {\boldsymbol{x}}_j \\
  \hat{\boldsymbol{a}}_{i_{2}} * {\boldsymbol{x}}_j \\
  \vdots \\
  \hat{\boldsymbol{a}}_{i_{\frac{N_{1}}{s}}} * {\boldsymbol{x}}_j
\end{bmatrix} \nonumber \\ 
&= \begin{bmatrix}
  1 &\dots & g_{i_{1}}^{\frac{N_{1}}{s} - 1} \\
  1 &\dots & g_{i_{2}}^{\frac{N_{2}}{s} - 1} \\
  \vdots & \ddots & \vdots \\
  1 & \dots & g_{i_{\frac{N_{1}}{s}}}^{\frac{N_{1}}{s} - 1} 
\end{bmatrix}^{-1}
\begin{bmatrix}
  \hat{\boldsymbol{a}}_{i_{1}} * {\boldsymbol{x}}_j \\
  \hat{\boldsymbol{a}}_{i_{2}} * {\boldsymbol{x}}_j \\
  \vdots \\
  \hat{\boldsymbol{a}}_{i_{\frac{N_{1}}{s}}} * {\boldsymbol{x}}_j
\end{bmatrix}
\end{align*}
where $\{i_{k}\}^{\frac{N_{1}}{s}}_{k=1}$ represent any $\frac{N_{1}}{s}$ distinct indices of $\{1, 2, ..., \frac{Ps}{N_{2}}\}$, and ${V_{j}}$ is a sub-matrix of   $V$. 
Finally, the master node can reconstruct $\boldsymbol{a}*\boldsymbol{x}$ after obtaining $\{\boldsymbol{a}*\boldsymbol{x}_{j}\}_{j=1}^{\frac{N_{2}}{s}}$. 

It should be noted that although this strategy can effectively reduce the straggler effect, it has the following limitations:
1) It cannot address the node heterogeneity and dynamic features of UAV-based NAC; 2) It is only resilient to up to $P - \frac{N_{1}N_{2}}{s^{2}}$ node failures; 3) In order to achieve high resilience, the introduced computation redundancy, indicated by $s - \sqrt{\frac{N_{1}N_{2}}{P}} > 0$,  should be large; 4) It also directly sends the input data to the worker nodes and thus may cause information leakage. 

\section{Dynamic Coded Convolution Strategy with Privacy Awareness}
\label{sec:method}
In this section, we introduce a privacy-aware dynamic coded convolution strategy that addresses the unique features of UAV-based NAC systems. How to decompose and encode the task is first explained, followed by the description of how to allocate and distribute the decomposed subtasks.  

\subsection{Task Decomposing and Encoding}
Instead of partitioning both vectors, we split only the input vector $\boldsymbol{x}$ evenly into $\frac{N_2}{b}$ sub-vectors 
$\{\boldsymbol{x}_{1}, \boldsymbol{x}_{2}, ..., \boldsymbol{x}_{\frac{N_2}{b}}\}$, where $b$ is the length of each sub-vector and can be any integer between 1 and $N_2$.  
Then instead of encoding $\boldsymbol{a}$, we encode the input sub-vectors into a larger set $\{\boldsymbol{\hat{x}}_{1}, \boldsymbol{\hat{x}}_{2}, ..., \boldsymbol{\hat{x}}_{\frac{N_2}{b}+k}\}$ by applying a ($\frac{N_{2}}{b}+k$, $\frac{N_{2}}{b}$) MDS code, 
where $k\in \mathbb{Z}^+$ specifies the computation redundancy. This will not only enhance the system resilience to uncertain stragglers, but also protect the data privacy to certain extent as the original input data is not sent. 
These sub-vectors $\{\boldsymbol{\hat{x}}_{i}\}_{i=1}^{\frac{N_2}{b}+k}$ are then pushed into a stack, denoted as $\boldsymbol{S}$, at the master node. Whenever a worker node becomes available, we pop a sub-vector $\hat{\boldsymbol{x}}_i$ from the top of stack $\boldsymbol{S}$ 
and send it 
to this worker node to compute $\boldsymbol{a} * \boldsymbol{\hat{x}}_{i}$, where vector $\boldsymbol{a}$ is pre-stored in all worker nodes.  Once the master node receives $\frac{N_2}{b}$ convolution results from the worker nodes, it can decode $\{{\boldsymbol{a}} * {\boldsymbol{x}}_i|1 \leq i \leq \frac{N_2}{b}\}$, using the similar decoding procedure described in Section \ref{coded_method}, and thereby reconstructing $\boldsymbol{a}*\boldsymbol{x}$. 

Unlike in the traditional coded convolution strategy, where the amount of computation redundancy is fixed after specifying the length $s$ of the sub-vectors, we here base on the network condition to dynamically introduce redundancy  when needed. In particular, we first set $k$ as a small value, e.g., $1$, so that the initial stack $\boldsymbol{S}$ only contains encoded input sub-vectors merely adequate enough for obtaining the final result. During task execution, whenever the stack $\boldsymbol{S}$ becomes empty (or below a certain threshold) and the master node still hasn't received sufficient results for computing the final value, the master node pushes a new  $\hat{\boldsymbol{x}}_i$, generated by encoding $\{\boldsymbol{x}_{i}\}_{i=1}^{\frac{N_2}{b}}$, into the stack. Note that we can pre-store an encoding matrix $V$ that is large enough at the master node, and take the first $\frac{N_2}{b}+k$ rows to initialize the stack $\boldsymbol{S}$ and take a new row whenever needed to generate new $\hat{\boldsymbol{x}}_i$ during task execution. With this scheme, we can  minimize the amount of introduced computation redundancy and maximize the system resilience to uncertain stragglers simultaneously.

\begin{figure}[b]
    \centering
    \includegraphics[width=0.45\textwidth]{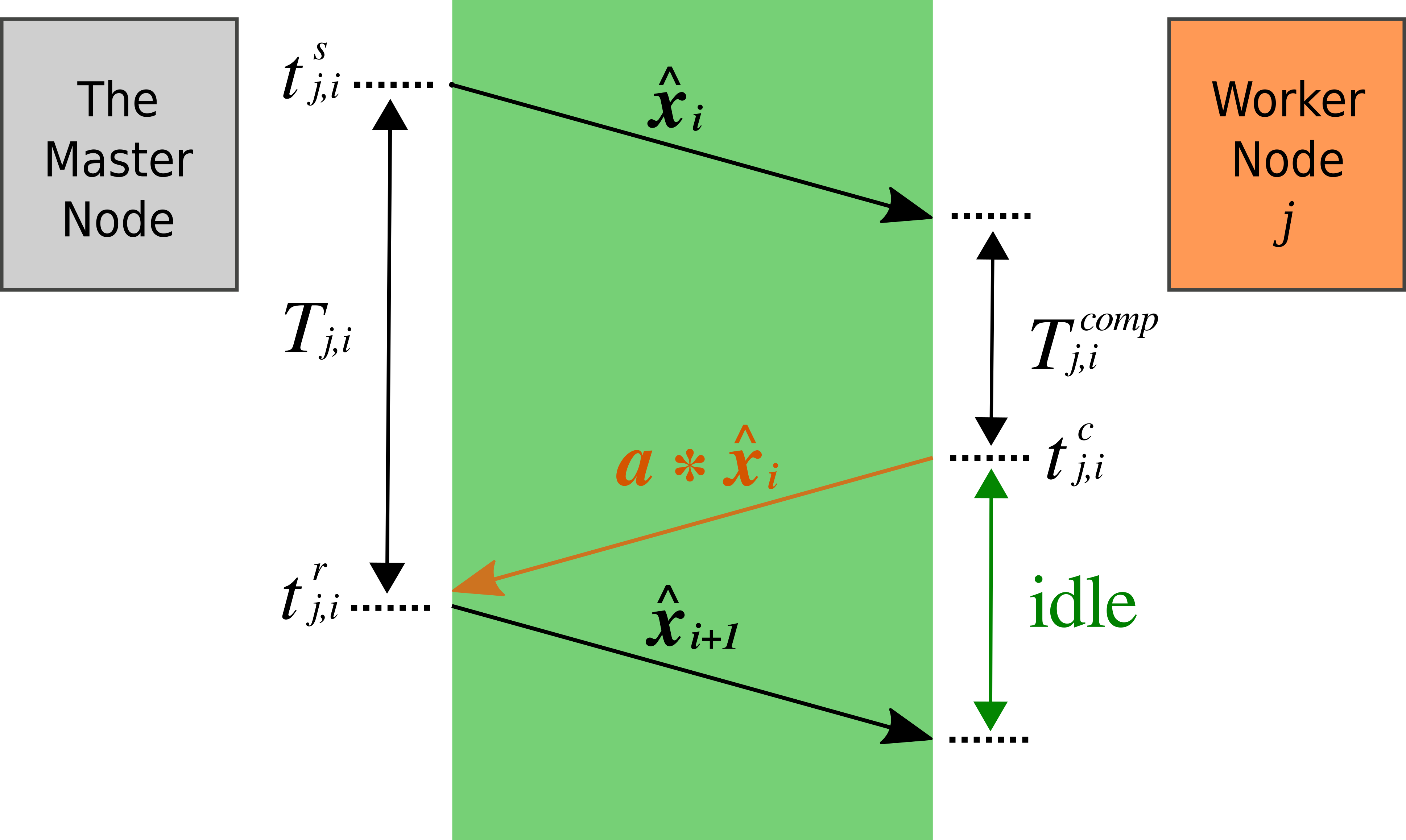}
    \caption{The communication flow between the master node and the worker node $j$.}
    \label{fig:illustration1}
\end{figure}
\subsection{Task Allocation}
To determine which worker node  the master node should send the next sub-vector $\hat{\boldsymbol{x}}_i$ (popped from the stack $\boldsymbol{S}$) to and when to send this sub-vector, we borrow the idea introduced in \cite{keshtkarjahromi2018dynamic}. The key idea is to send a sub-vector $\hat{\boldsymbol{x}}_i$ to each worker node at the beginning. The master node then determines the best time to send the next sub-vector to a worker node based on the estimation of the time required for this worker node to complete its current task as well as send back the result. 

In particular, let $\mathcal{T}_{j,i}$ be the time interval between sending two consecutive sub-vectors, $\boldsymbol{\hat{x}}_{i}$ and $\boldsymbol{\hat{x}}_{i+1}$, to the worker node $j$ from the master node. As illustrated in Fig. \ref{fig:illustration1}, in order to maximally reduce the computation delay, the desired $\mathcal{T}_{j,i}$ will minimize the idle time at the worker node $j$ while not overloading it. That is, ideally, the worker node $j$ should receive $\boldsymbol{\hat{x}}_{i+1}$ immediately after it completes the previous task, i.e., computing $\boldsymbol{a} * \boldsymbol{\hat{x}}_{i}$. To determine $\mathcal{T}_{j,i}$, the key is thus to estimate the time required for the master node to compute $\boldsymbol{a} * \boldsymbol{\hat{x}}_{i}$, denoted as $T^{comp}_{j,i}$. Here, we  apply the method introduced in \cite{keshtkarjahromi2018dynamic} to estimate the expected time required for the worker node $j$ to compute $\boldsymbol{a} * \boldsymbol{\hat{x}}_{i}$.  In particular, the expected computation time $\mathbb{E}[T^{comp}_{j,i}]$ can be estimated by following equations: 
\begin{equation}
\mathbb{E}[T^{comp}_{j,i}] \approx \frac{t_{j,i}^{c} - t_{j,i}^{u}}{c_j}  \label{eq:eq1}
\end{equation}
\begin{equation}
t_{j,i}^{c} \approx t_{j,i}^{r} - \frac{B_r}{B_r + B_x}RTT_j
\end{equation}
\begin{equation}
t_{j,i}^{u} \approx t_{j,i-1}^{u} + \max(0, RTT_j - t_{j,i-1}^{r} - t_{j,i}^{s})
\end{equation}
where $t_{j,i}^{c}$ is the time when the worker node $j$ finishes computing $\boldsymbol{a} * \boldsymbol{\hat{x}}_{i}$, $t^{r}_{j,i}$ is the time when the master node receives the computation result of $\boldsymbol{a} * \boldsymbol{\hat{x}}_{i}$ from the worker node $j$, and $t^{s}_{j,i}$ is the time when the master node sends sub-vector $\hat{\boldsymbol{x}}_i$ to the worker node $j$. $t_{j,i}^{u}$ is the accumulated idle time of worker node $j$. $c_j$ is the number of results sent back from the worker node $j$, $B_x$ (bytes) is the size of the vector $\boldsymbol{\hat{x}}_{i}$ and $B_r$ (bytes) is the size of the result of $\boldsymbol{a} * \boldsymbol{\hat{x}}_{i}$. Lastly, $RTT_j$ is the round trip time of sending $\boldsymbol{\hat{x}}_{i}$ to the worker node $j$ and receiving the computed result, which can be estimated at the master node by exchanging Acknowledgement (ACK) packages \cite{sessini2006observations} or based on the timestamps returned by the worker node $j$.

Given $
\mathbb{E}[T^{comp}_{j,i}]$, we then determine $\mathcal{T}_{j,i}$ using the following equation:
\begin{equation} \label{eq:4}
\mathcal{T}_{j,i} = \min(t^r_{j,i} - t^s_{j,i}, \mathbb{E}[T^{comp}_{j,i}])
\end{equation}
Algorithm \ref{alg:2} summarizes the complete procedure of the proposed dynamic coded convolution strategy. 

\begin{algorithm}
\caption{Dynamic Coded Distributed Convolution Strategy}\label{alg:2}
\KwData{$\boldsymbol{a}$, $\boldsymbol{x}$, $b$, $V$}
\KwResult{$\boldsymbol{a} * \boldsymbol{x}$}
$k \gets 1$, $N_{1} \gets |\boldsymbol{a}|$, $N_{2} \gets |\boldsymbol{x}|$\; 
Partition $\boldsymbol{x}$ into a set of sub-vectors $\{\boldsymbol{x}_{i}\}_{i=1}^{\frac{N_2}{b}}$ with each of length $b$\;
Use the first $\frac{N_2}{b}+k$ rows of $V$ to encode $\{\boldsymbol{x}_{i}\}_{i=1}^{\frac{N_2}{b}}$ into a larger set $\{\boldsymbol{\hat{x}}_{i}\}_{i=1}^{\frac{N_2}{b}+k}$, and push them into the stack $\boldsymbol{S}$\;
$\mathcal{R} \gets$ empty stack for storing received results\;
$\mathcal{P} \gets$ list of worker nodes within master node's communication range\; 
\For{\text{each node} $j$ in $\mathcal{P}$}{
  Send $\boldsymbol{\hat{x}}_{i}$ popped from $\boldsymbol{S}$ to node  $j$\;
  $t^s_{j,i} \gets current\_time()$\;
}
\EndFor
\While{$|\mathcal{R}| < \frac{N_{2}}{b}$}{
  \For{\text{each node} $j$ in $\mathcal{P}$}{
    \If{$|\boldsymbol{S}| \leq 1$}{
     $k \gets k + 1$\;
   Use the $k$-th row in $V$ to generate 
      $\boldsymbol{\hat{x}}_{\frac{N_2}{b}+k}$ and then push it into $\boldsymbol{S}$\; 
    }
    \If{$current\_time() \geq t^s_{j,i} + \mathcal{T}_{j,i}$ }{ 
      Send $\boldsymbol{\hat{x}}_{i+1}$ popped from  $\boldsymbol{S}$ to node $j$\;
     $t^s_{j,i+1} \gets current\_time()$\;
    }
    \If{receiving result of $\boldsymbol{a}*\hat{\boldsymbol{x}}_i$ from node $j$}{
      Push the result into  $\mathcal{R}$\;
      Update $\mathcal{T}_{j,i}$ using \eqref{eq:eq1}-\eqref{eq:4}\;
    }
  }\EndFor
}
$\text{Reconstructs } \boldsymbol{a} * \boldsymbol{x} \text{ using results from } \mathcal{R}$\;
\end{algorithm}

\section{Simulation Studies}
\label{sec:simulation}

In this section, we conduct simulations to evaluate the performance of the proposed strategy, in comparison with the uncoded convolution and traditional coded convolution schemes. All simulations are performed on a PC with 16GB of RAM and Intel Core i5-4590.

\subsection{System Models} 
We use the following system models to simulate the movement of UAVs, as well as how they compute and how they communicate with each other. 

\subsubsection{Mobility Model}
A simple 2-dimensional (2D) point-mass mobility model is adopted to simulate the movement of each UAV. In particular, let $\boldsymbol{p}_j(t)$ denote the location of UAV $j$ at time $t$. Then its location at the next time point $t'$ is given by the following equation: 
\begin{equation}
\boldsymbol{p}_{j}(t') = \boldsymbol{p}_{j}(t) + \boldsymbol{v}_{j}(t)(t' - t) \nonumber
\end{equation}
where $\boldsymbol{v}_{j}(t)$ is the velocity of UAV $j$ at time $t$, where $t' \geq t$.


\subsubsection{Computing Model}

To simulate the time required by each UAV $j$ to convolve two vectors, say $\boldsymbol{e}_1 \in \mathbb{R}^{n_1}$ and $\boldsymbol{e}_2 \in \mathbb{R}^{n_1}$, 
we adopt the following shifted exponential distribution model commonly used in the literature \cite{wang2021batch}:
\begin{equation}
    \mathrm{Pr}[T^{comp}_{j}(\boldsymbol{e}_1, \boldsymbol{e}_{2}) \leq t] = 1 - e^{\frac{\mu_{j}}{c(n_1,n_2)}(t - \alpha_{j}c(n_1,n_2))} \nonumber
\end{equation}
where $T^{comp}_{j}(\boldsymbol{e}_1, \boldsymbol{e}_{2})$ is the time taken by UAV $j$ to compute $\boldsymbol{e}_1* \boldsymbol{e}_{2}$. $\alpha_{j} > 0$ and $\mu_{j} > 0$ are shift and straggling parameters, respectively, which characterize the computing power of UAV $j$.
$c(n_1, n_2)$ represents the computation load required for computing $\boldsymbol{e}_1* \boldsymbol{e}_{2}$.  
To derive this value, we assume that the Fast Fourier Transform (FFT) is used by each UAV  to calculate the convolution of two vectors with arbitrary lengths. Hence, we have \cite{cooley1969fast,bracewell1986fourier}:
\begin{align}
c(n_1, n_2)
 & = \mathcal{O} \big((n_{1} + n_{2} - 1)(\log(n_{1} + n_{2} - 1) + 1)\big) \nonumber\\
 & = C(n_{1} + n_{2}) \log(n_{1} + n_{2}) \nonumber
\end{align}
where $C$ is a constant independent of the lengths of the vectors. 


\subsubsection{Communication Model}
We assume that the communication between any two UAV nodes is achieved through a directional antenna, and the antennas are always aligned during the movement \cite{chen2017long,li2019design}. The communication time required for the master node to transmit to (or receive from) a worker node $j$ a dataset containing $n$ numbers at time $t$ can be approximated by the following equation \cite{liu2019learning}: 
\begin{equation}
T^{comm}_{j}(n,t) = \frac{n \times u}{R_{j}(t)} \nonumber
\end{equation}
where $u$ is the average size of the numbers in the dataset. $R_{j}(t)$ is the data rate (bits/sec) given by:
\begin{equation}
R_{j}(t) = B\log_{2}(1 + \frac{10^{\frac{S_{d}(t) - 30}{10}}}{N_0}) \nonumber
\end{equation}
where $B$ (Hz) is the communication bandwidth between the master node and worker node $j$ and 
$N_0$ is the noise power (W), both of which are assumed to be constant. $S_{d}(t) = P_t + 20log_{10}(\lambda) -20log_{10}(4\pi)-20log_{10}(d(t))+G_{l|dBi}+w$ is the signal power (W) that depends on the distance $d(t)$ between the master node and the worker node $j$ at time $t$. $P_t$ (dBm) is the transmitting power, $G_{l|dBi}$ is the sum of the transmitting and receiving gains, $w$ is the Gaussian noise and $\lambda$ is the wave length. 


\subsection{Experiment Setup}
We consider the following four computation scenarios: 
\begin{itemize}
  \item \textbf{Scenario 1:} $N_{1} = 2^{12}$, $N_{2} = 2^{11}$, $P = 8$.
  \item \textbf{Scenario 2:} $N_{1} = 2^{12}$, $N_{2} = 2^{11}$,  $P = 4$.
  \item \textbf{Scenario 3:} $N_{1} = 20000$, $N_{2} = 30000$, $P = 8$.
  \item \textbf{Scenario 4:} $N_{1} = 20000$, $N_{2} = 30000$, $P = 6$.
\end{itemize}
In all scenarios, the straggling parameter $\mu_{j}$ in the computation model is randomly sampled from the range $[3 \times 10^{6}, 6 \times 10^{6}]$, and the shift parameter $\alpha_{j}$ is set to $\alpha_{j} = \frac{1}{\mu_{j}}$. 
To simulate the straggler effect, we consider two cases: 1) stragglers caused by long communication latency and/or computation delay; and 2) stragglers caused by system failures or  moving out of the master node's communication range. To model the first case, we manually 
make the run-time of the stragglers 
to be $15$ times the simulated run-time returned from its computation model. To model the second case, we make the stragglers stop returning any results to the master node. 

For the configuration of the mobility model, the initial position of each UAV is randomly sampled from the range $[(-1500, -1500), (1500, 1500)]$. The velocity of each UAV is randomly sampled from the range $[(-10, -10), (10, 10)]$ m/s once every second. 
Lastly, the parameters in the communication model are configured as $B=10^{6}$, $N_0=10^{-12}$, and $S_{d}(t)=6-20log_{10}(d(t))$. It is worthy of remark that our method does not require any knowledge of the mobility, computation or communication models. 
\begin{figure}[t]
    \centering
    \includegraphics[width=0.45\textwidth]{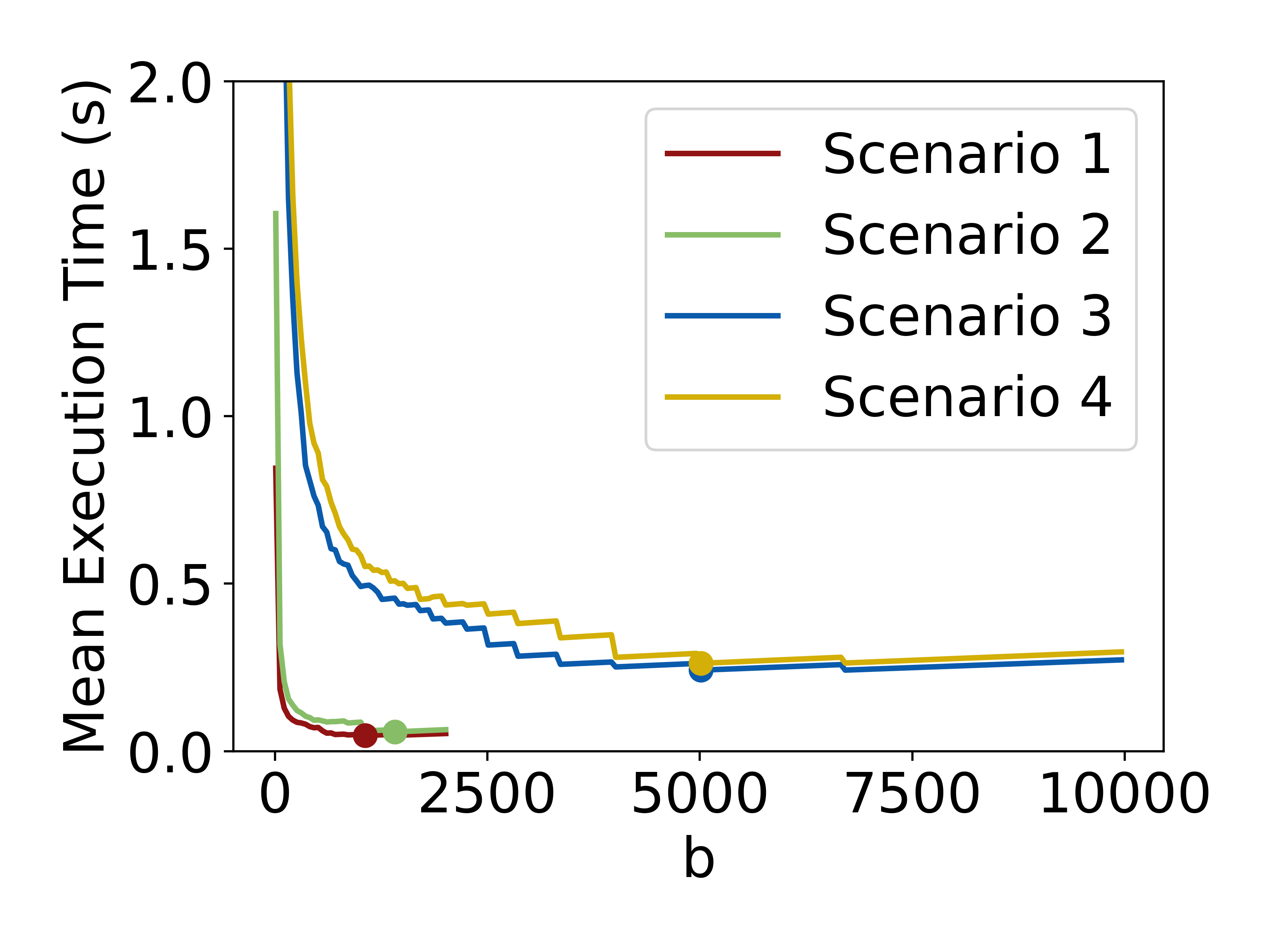}
    \caption{Impact of parameter $b$ on the performance of the proposed strategy.}
    \label{fig:cp}
\end{figure}
\subsection{Simulation Results}
\subsubsection{Impact of Parameter $b$} We first study the impact of the key parameter in our method, i.e., the length of the input sub-vectors $b$, by evaluating the performance of our method at different values of $b$. To reduce uncertainty, each experiment in our simulation study is repeated for $25$ times and the mean execution times are recorded.  
As shown in Fig. \ref{fig:cp}, in all four scenarios, the execution time of our method first decreases as $b$ increases, and then increases after $b$ reaches a certain value. The best $b$ is thus the one that leads to the minimum mean execution time, which varies in different scenarios. 
Fig. \ref{fig:cp} also reveals that with the increase of the problem size (characterized by $N_1$ and $N_2$) or the decrease of the amount of computing resources available (indicated by $P$), the time required for conducting the computation task increases. 

In subsequent experiments, we use the best-performing $b$ to configure our method in each scenario, which are marked with big dots in Fig. \ref{fig:cp}. 
For the choice of $s$, length of the sub-vectors in the traditional coded convolution strategy, we follow the selection guideline provided in \cite{dutta2017coded} and set $s$ as the integer from range $[\sqrt{\frac{N_{1}N_{2}}{P}},\min({N_{1},N{2})}]$ that maximizes $|\epsilon(s)|$, where $\epsilon(s)$ is given by:
\begin{equation}
\epsilon(s) = -\sum_{j=1}^{P} \frac{
  (\frac{Ps}{N_2} - \frac{N_1}{s} + 1)\mu_{j}^{\alpha_{j}}  
}{
  P((2Cs)\log(2s))^{\alpha_{j}}
} \nonumber
\end{equation}

\begin{figure}[b]
    \centering
    \includegraphics[width=0.45\textwidth]{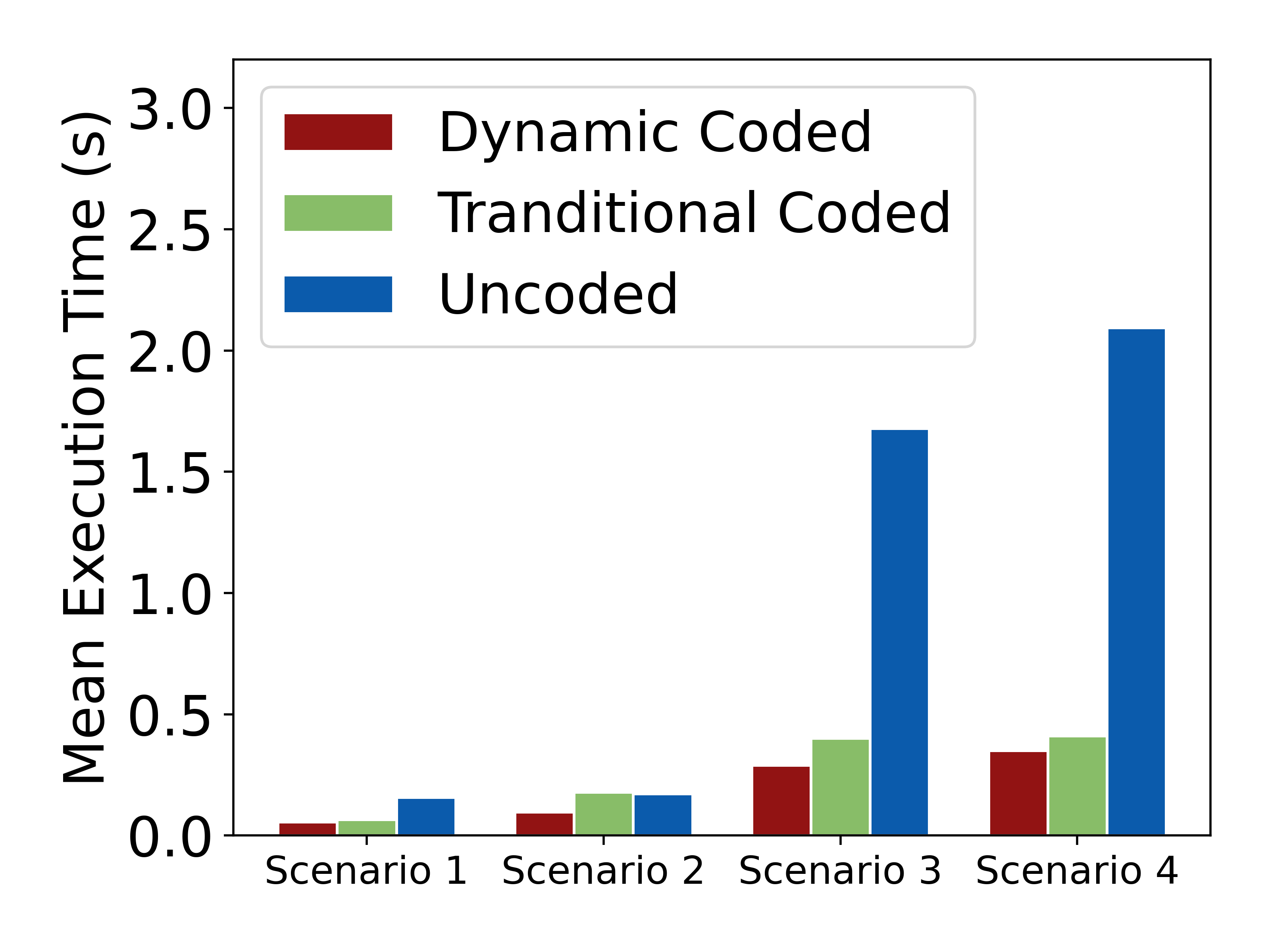}
    \caption{Comparison of different strategies when 50\% of the worker nodes are stragglers with long communication/computation delays.}
    \label{fig:comp}
\end{figure}

\begin{figure}[t]
    \centering
    \includegraphics[width=0.45\textwidth]{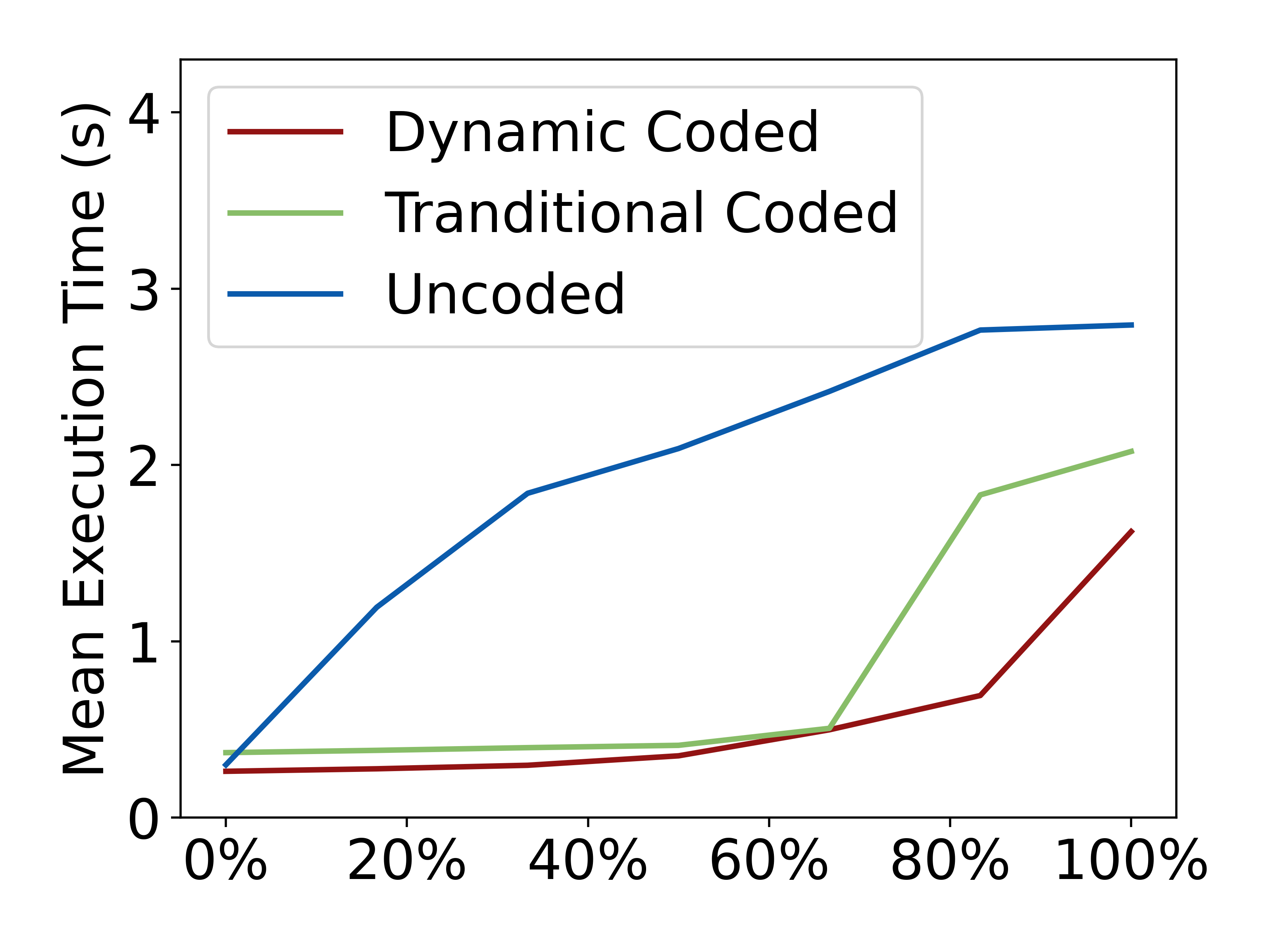}
    \caption{Comparison of different strategies in Scenario 4 when the straggler ratio increases.}
    \label{fig:stress}
\end{figure}

\begin{figure}[tb]
    \centering
    \includegraphics[width=0.45\textwidth]{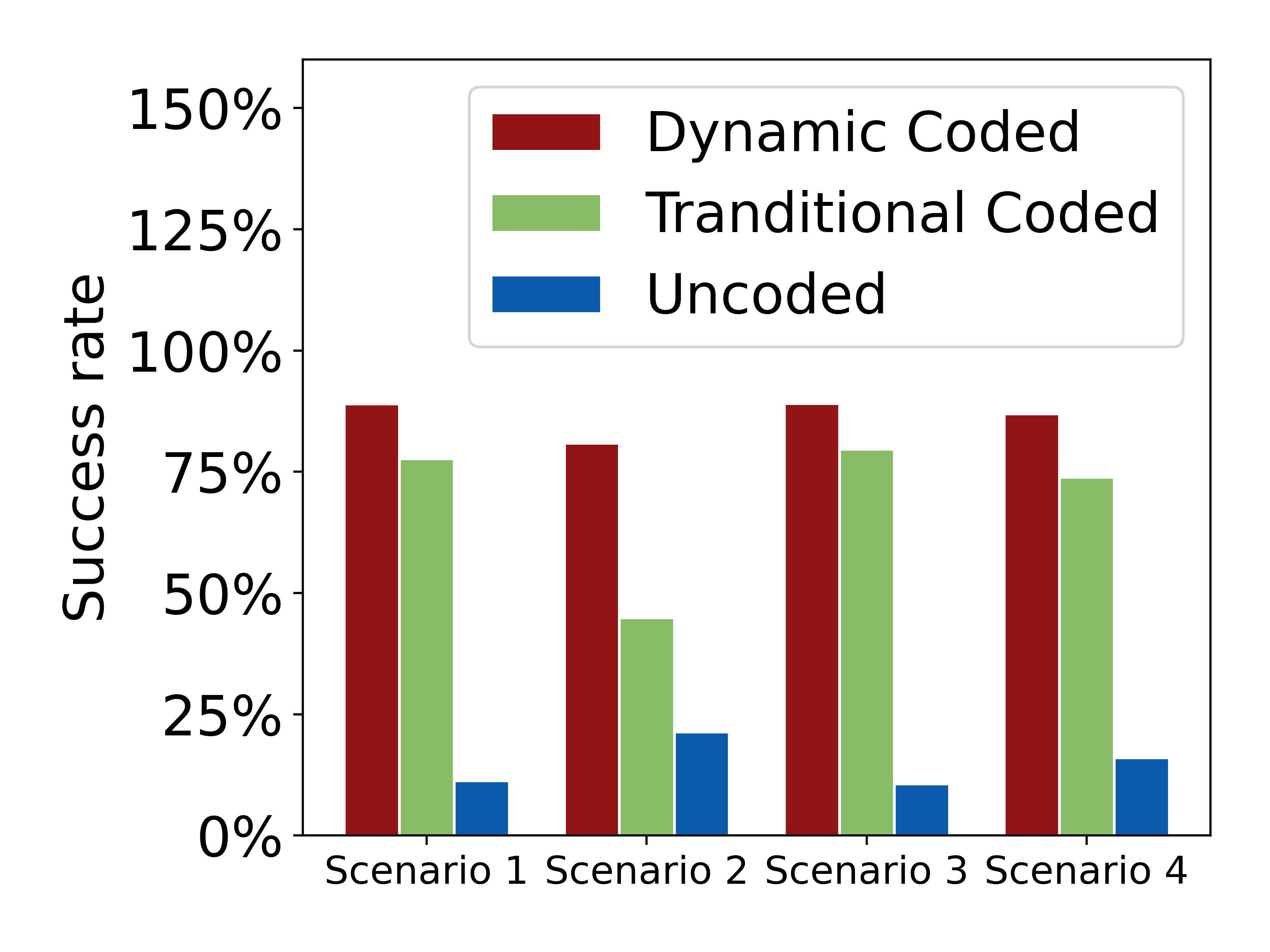}
    \caption{The task success rate of different strategies when node failures/leaves can happen.}
    \label{fig:cp3}
\end{figure}
\subsubsection{Comparison Studies}
We first study the case when stragglers with long communication/computation delay are present. 
Fig. \ref{fig:comp} compares the performance of different strategies when 50\% of the worker nodes randomly selected are such stragglers. It shows that our method achieves the highest efficiency in all scenarios and the uncoded convolution strategy is the least efficient. 

To better understand the three strategies, we further conduct a stress test by varying the percentages of the stragglers with long delays. Fig. \ref{fig:stress} shows the results of the stress test for different strategies in Scenario 4. As we can see, the performance of all three strategies degrade with the increase of the straggler ratio and our method achieves the best performance in all cases. It can also be observed that the uncoded strategy is the most sensitive to stragglers, as indicated by the immediate increase of its execution time when the straggler ratio becomes non-zero. Nevertheless, the execution time of our method does not increase much until the straggler ratio exceeds around 83\%, demonstrating its high resilience to uncertain stragglers.   

Lastly, we investigate the case when node failures or node leaves can happen. As the master node cannot receive any results from such stragglers, the results received from other worker nodes may not be sufficient enough for the master node to reconstruct the convolution $\boldsymbol{a} * \boldsymbol{x}$, leading to task failures. Therefore, in this study, we measure the task success rate (ratio of successful runs) of each strategy at the presence of such type of stragglers. 
Fig. \ref{fig:cp3} shows the success rates of different strategies, where each strategy runs 2000 times in each scenario. In each simulation run, $0$ up to $P$ worker nodes can fail. 
The result demonstrates that our method is highly resilient to node failures. It is worthy noting that our method can successfully complete the task as long as there is a worker node alive, which can be the master node itself. Moreover, even if the master node loses connection with all worker nodes, as long as there is a new node joining later, the task will resume.



\section{Conclusion}
\label{sec:conclusion}
This paper introduces an efficient, resilient, and privacy-aware distributed computing strategy for vector convolution tasks in heterogeneous and mobile UAV-based NAC systems. It combines the coding theory with a novel task decomposing and allocation mechanism to achieve a high resilience to uncertain stragglers with the minimal computation redundancy. As input data is encoded, it also provides some protection for data privacy. 
The simulation results show that the proposed strategy outperforms existing solutions in both efficiency and resilience, especially when a large number of high-latency computing nodes are present or frequent node leaves/failures  happen. Moreover, our method is adaptive to the dynamic network changes in UAV-based NAC systems and can complete the task as long as there is a worker node alive, which can be the master node itself. 

In the future, we will design intelligent strategies to automate the configuration for the key parameter $b$, and extend the proposed strategy to real CNN-based applications. We will also develop  hardware testbed for UAV-based NAC and conduct flight tests to evaluate the performance of the proposed strategy.  

\section*{Acknowledgment}
We would like to thank the National Science Foundation (NSF) under Grants CI-1953048 and CAREER-2048266  for the support of this work.



%
\begin{small}
	\bibliographystyle{IEEEtran}
	\bibliography{IEEEabrv, reference}
\end{small}

\end{document}